\documentclass[aps,showpacs,tightenlines,twocolumn,nofootinbib,nobibnotes,superscriptaddress]{revtex4-1}
\usepackage{amsmath,amssymb,amsfonts,bm}
\usepackage{graphicx}
\usepackage{epstopdf}
\usepackage{dcolumn}
\usepackage{mathrsfs}
\usepackage{tgtermes}
\usepackage[colorlinks=true,linkcolor=red,citecolor=blue, urlcolor=blue,bookmarks=false]{hyperref}
\bibpunct{[}{]}{,}{n}{}{}
\usepackage{verbatim}
\usepackage{amsthm,amsmath,amssymb}
\usepackage{mathrsfs}
\usepackage{booktabs}
\usepackage{multirow}

\begin{document}
\title{High second Chern number induced by long-range hopping in a four-dimensional Dirac model}
\date{\today}
\author{Zheng-Rong Liu}
\affiliation{Department of Physics, Hubei University, Wuhan 430062, China}
\author{Xiang Liu}
\affiliation{Department of Physics, Hubei University, Wuhan 430062, China}
\author{Rui Chen}\email{chenr@hubu.edu.cn}
\affiliation{Department of Physics, Hubei University, Wuhan 430062, China}
\author{Bin Zhou}\email{binzhou@hubu.edu.cn}
\affiliation{Department of Physics, Hubei University, Wuhan 430062, China}
\affiliation{Key Laboratory of Intelligent Sensing System and Security of Ministry of Education, Hubei University, Wuhan 430062, China}
\affiliation{Wuhan Institute of Quantum Technology, Wuhan 430206, China}

\begin{abstract}
Four-dimensional (4D) topological systems provide a promising platform for exploring topological phenomena beyond three dimensions. So far, extensive recent studies on 4D topological insulators have focused on the 4D Dirac model, while its second Chern number is restricted to a limited set of values.
In this work, we demonstrate that introducing long-range hopping into the 4D Dirac model induces topological phases with high second Chern numbers. Furthermore, we show that the long-range hopping can transform a trivial insulator into a topological insulator with a nonzero second Chern number.
Our work establishes long-range hopping as a powerful route for engineering 4D topological states and reveals new possibilities for realizing unconventional topological phases beyond minimal models.
\end{abstract}

\maketitle

\section{Introduction}
Topological phases of matter have attracted considerable interest due to their quantized boundary transport and unconventional bulk topological responses~\cite{RevModPhys.82.3045, RevModPhys.83.1057, Bernevig_2013, RevModPhys.88.021004, RevModPhys.88.035005, Shen_2017, RevModPhys.89.040502, RevModPhys.89.041004, W_lfle_2018}.
Beyond conventional three-dimensional systems, higher-dimensional topological systems provide a broader framework for exploring novel topological phenomena associated with higher-dimensional topological invariants~\cite{PhysRevB.78.195424, doi:10.1126/science.aam9031, PhysRevB.98.094434, PhysRevB.98.125431}.
In particular, four-dimensional (4D) topological insulators characterized by the second Chern number $C_2$ exhibit unique topological properties, including a quantized nonlinear electromagnetic response and a bulk-boundary correspondence between nontrivial bulk topology and three-dimensional gapless boundary states~\cite{10.1126/science.294.5543.823}.
These features reveal topological structures that have no direct counterparts in lower-dimensional systems~\cite{10.1126/science.294.5543.823, doi:10.1126/science.aam9031, PhysRevLett.111.186803}.
Although a genuine 4D space is inaccessible in conventional materials, synthetic dimensions~\cite{PhysRevA.93.043827, 10.1093/nsr/nwac289, PhysRevX.13.011003, PhysRevLett.111.226401} and artificial quantum systems~\cite{PhysRevLett.108.133001, PRXQuantum.2.010310} have enabled experimental investigations of 4D topological phases through ultracold atoms~\cite{PhysRevLett.115.195303, 10.1038/nature25000}, photonic lattices~\cite{10.1038/nature25011, PhysRevA.87.013814, 10.1515/nanoph-2022-0778, PhysRevB.101.205141}, acoustic systems~\cite{PhysRevX.11.011016}, and electric circuits~\cite{10.1038/s41467-020-15940-3, 10.1093/nsr/nwaa065, 10.1038/s41467-023-36359-6, 10.1038/s41467-023-36767-8}.

The 4D minimal Dirac model serves as a paradigmatic platform for studying 4D topological phenomena~\cite{PhysRevLett.109.135701, PhysRevLett.129.196602, PhysRevResearch.2.023364, PhysRevB.108.085306, PhysRevB.110.195144, 10.1038/s41467-023-36767-8}. 
Extensive studies of 4D topological insulators have focused on this minimal model~\cite{PhysRevB.78.195424, 10.1209/0295-5075/ad397c, PhysRevB.97.134303, PhysRevB.101.205141, PhysRevLett.109.135701, PhysRevLett.129.196602, PhysRevResearch.2.023364, PhysRevB.108.085306, PhysRevB.109.125303, 10.1088/0256-307X/41/4/047102}, where the topological nontrivial phases are characterized by second Chern numbers of $C_2=\pm1$ and $\pm3$. 
Recent works have demonstrated that additional mechanisms, such as external magnetic fields and periodic driving, can further enrich the topological phase diagram and enable phases with even second Chern numbers~\cite{PhysRevB.109.125303, 10.1088/0256-307X/41/4/047102, 10.1088/1402-4896/ae8e4d}. 
However, these approaches rely on external controls, and how intrinsic modifications to the lattice Hamiltonian, such as long-range hopping, expand the topological phase space of higher-dimensional systems remains unexplored.

Long-range hopping provides an effective route to modify band structures and generate unconventional topological phases in low-dimensional systems~\cite{10.1088/1367-2630/aa84d0, PhysRevB.93.014517, PhysRevB.100.235452}.
In contrast to conventional nearest-neighbor hopping models, long-range hopping introduces additional momentum-dependent terms and enables higher winding numbers~\cite{PhysRevB.109.035114, PhysRevA.106.012211, 10.1088/1402-4896/ae0fca, PhysRevB.111.014109, arXiv2601.14769, PhysRevB.111.104204, PhysRevApplied.19.054028, 10.1103/ctfq-jsj1, arXiv2604.09801}, multiple edge states~\cite{PhysRevB.111.024507, PhysRevB.108.184303, PhysRevB.88.165111}, and topological phases beyond nearest-neighbor hopping models~\cite{PhysRevLett.131.186303, PhysRevLett.123.025301, PhysRevLett.119.023001, PhysRevLett.113.156402, 10.1002/apxr.202500240, PhysRevB.83.075105, PhysRevB.93.041102, PhysRevB.93.125128, PhysRevB.94.125121, PhysRevB.99.035146, 10.1016/j.aop.2016.07.026}.
Recent theoretical and experimental studies have demonstrated that long-range hopping can significantly enrich topological phenomena in one- and two-dimensional systems, including long-range Su-Schrieffer-Heeger models~\cite{PhysRevB.97.064304, PhysRevA.106.012211, 10.1088/1402-4896/ae0fca, PhysRevB.111.014109, 10.1088/1572-9494/ac75db, 10.1088/1402-4896/ae2e58, 10.1038/s41534-019-0159-6, 10.1103/ctfq-jsj1, arXiv2412.19508}, Kitaev chains~\cite{PhysRevLett.113.156402, PhysRevLett.119.110601, PhysRevB.94.125121, PhysRevB.95.195160, PhysRevB.97.064304, PhysRevB.104.075113, 10.1103/mvcm-fbjl, 10.1007/JHEP05(2023)066}, and artificial higher-order topological insulators with long-range couplings~\cite{PhysRevLett.128.127601, PhysRevApplied.23.024024, PhysRevB.108.205135, arXiv2512.23168}.
These developments raise an intriguing question: whether long-range hopping can similarly overcome the limitations of the 4D minimal Dirac model and generate unexplored topological phases with high second Chern numbers.

In this work, we address this question by investigating the effects of long-range hopping on 4D Dirac systems.
By introducing the next-nearest-neighbor and next-next-nearest-neighbor hoppings, we demonstrate that the long-range hopping provides a mechanism for generating and controlling 4D topological phases with high second Chern numbers.
Starting from a 4D trivial insulator, we show that the long-range hopping induces topological phase transitions into phases with high $|C_2|$, including $C_2=-6$.
Moreover, when applied to the 4D topological insulators, the long-range hopping enables transitions into new topological phases with high second Chern numbers, such as $C_2=-6$ and $C_2=-7$, which are absent in the 4D minimal Dirac model.
The corresponding gapless boundary states verify the bulk-boundary correspondence of these high second Chern number phases.
Our work demonstrates that the long-range hopping can expand the topological phase space of the 4D minimal Dirac model and induces the high second Chern number phases absent in the original model.

The rest of this paper is organized as follows. 
In Sec.~\ref{SecII}, we introduce the next-nearest-neighbor and next-next-nearest-neighbor hoppings into the 4D Dirac model and describe the method used to calculate the second Chern number. 
In Sec.~\ref{SecIII}, we investigate the effects of long-range hopping on 4D trivial insulators and demonstrate the emergence of topological phases with nonzero second Chern numbers. 
In Sec.~\ref{SecIV}, we explore how the long-range hopping modifies 4D topological insulators with nonzero second Chern numbers and leads to additional topological phases. 
Finally, Sec.~\ref{Conclusion} summarizes our main results and conclusions.

\begin{figure}[t]
	\includegraphics[width=0.48\textwidth]{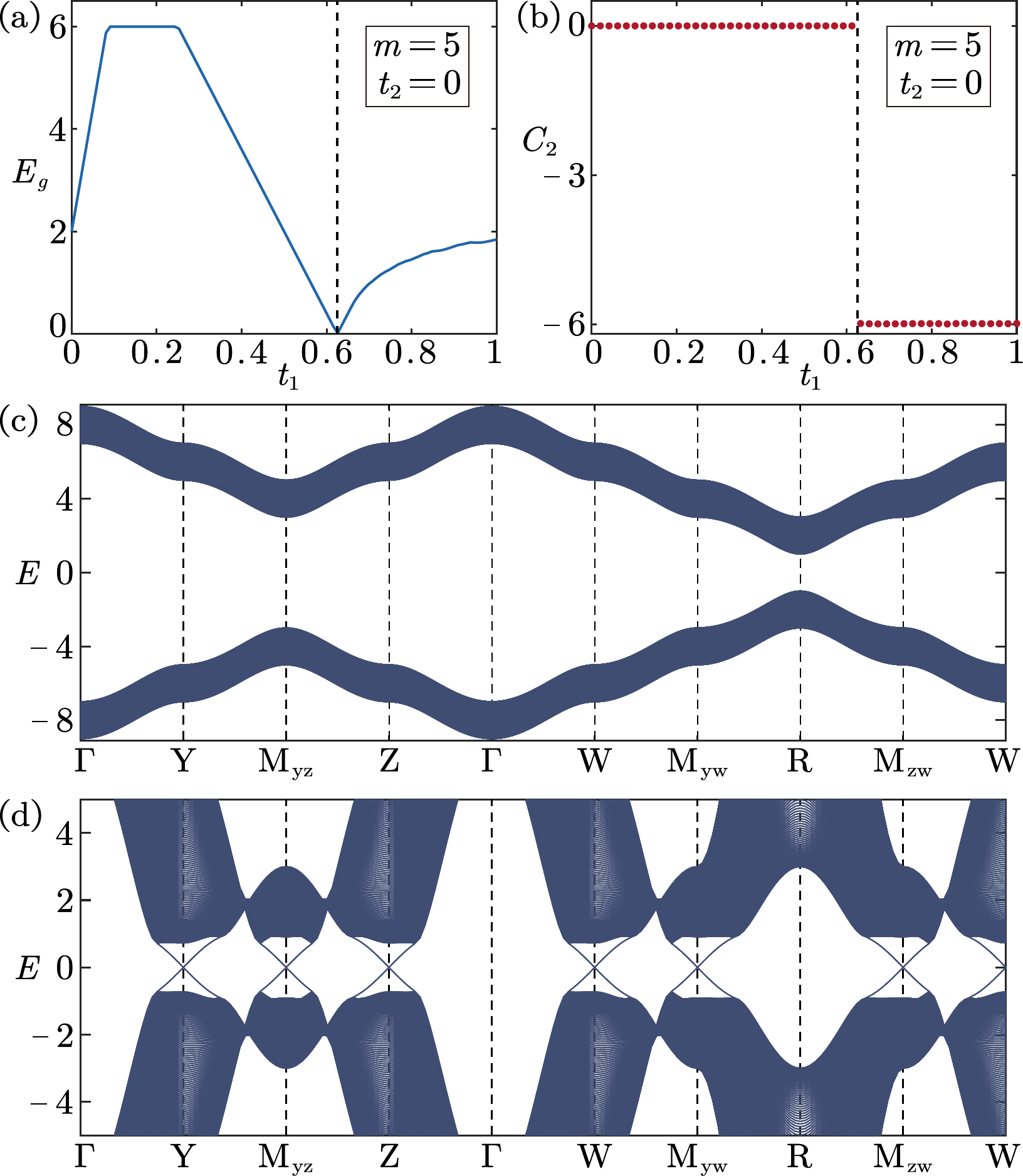} \caption{(a) Bulk energy gap $E_g$ as a function of the next-nearest-neighbor hopping amplitude $t_1$. The dashed line indicates the gap-closing point at $t_1=m/8$. (b) Second Chern number $C_2$ as a function of $t_1$. (c), (d) Energy spectra in the three-dimensional Brillouin zone under open boundary conditions along the $x$ direction for $t_1=0$ and $t_1=0.8$, respectively. The indicators of the horizontal axis are the high symmetry $\textbf{k}(k_{y}, k_{z}, k_{w})$ points in the three-dimensional Brillouin zone, $\Gamma(0, 0, 0)$, ${\rm{Y}}(\pi, 0, 0)$, ${\rm{Z}}(0, \pi, 0)$, ${\rm{W}}(0, 0, \pi)$, ${\rm{M_{yz}}}(\pi, \pi, 0)$, ${\rm{M_{yw}}}(\pi, 0, \pi)$, ${\rm{M_{zw}}}(0, \pi, \pi)$, ${\rm{R}}(\pi, \pi, \pi)$. Here, $m=5$ and $t_2=0$.}%
	\label{fig1}
\end{figure}

\section{4D Model with Long-Range Hopping}
\label{SecII}
In this section, we introduce the long-range hopping into the 4D Dirac model~\cite{PhysRevB.78.195424}. The 4D Hamiltonian with next-nearest-neighbor and next-next-nearest-neighbor hopping is given by
\begin{align}
	H(\mathbf{k})=&\sin(k_{x})\Gamma_{2}+\sin(k_{y})\Gamma_{3}+\sin(k_{z})\Gamma_{4}+\sin(k_{w})\Gamma_{5}\nonumber\\
	&+M(\mathbf{k})\Gamma_{1},
\end{align}
where the Dirac matrices $\Gamma_{j}=(\sigma_{x}\otimes \sigma_{0}, \sigma_{y}\otimes \sigma_{0}, \sigma_{z}\otimes \sigma_{x}, \sigma_{z}\otimes \sigma_{y}, \sigma_{z}\otimes \sigma_{z})$, $j=1, 2, 3, 4, 5$, satisfy the anticommutation relations $\{\Gamma_{i},\Gamma_{j}\}=2\delta_{ij}$. The last term in $H(\mathbf{k})$ is $M(\mathbf{k})=m+t_{0}[\cos(k_{x})+\cos(k_{y})+\cos(k_{z})+\cos(k_{w})]+t_{\rm{NNN}}(\mathbf{k})+t_{\rm{NNNN}}(\mathbf{k})$, where $m$ is the Dirac mass and $t_{0}$ is the nearest-neighbour hopping amplitude. In the subsequent calculations, we set $t_{0}=1$. The next-nearest-neighbor hopping term is
\begin{align}
	t_{\rm{NNN}}(\mathbf{k})=&4t_{1}[ \cos(k_{x})\cos(k_{y})+\cos(k_{x})\cos(k_{z})\nonumber\\
	&+\cos(k_{x})\cos(k_{w})+\cos(k_{y})\cos(k_{z})\nonumber\\
	&+\cos(k_{y})\cos(k_{w})+\cos(k_{z})\cos(k_{w}) ],
\end{align}
and the next-next-nearest-neighbor hopping term is
\begin{align}
	t_{\rm{NNNN}}(\mathbf{k})=&8t_{2}[ \cos(k_{x})\cos(k_{y})\cos(k_{z})\nonumber\\
	&+\cos(k_{x})\cos(k_{y})\cos(k_{w})\nonumber\\
	&+\cos(k_{x})\cos(k_{z})\cos(k_{w})\nonumber\\
	&+\cos(k_{y})\cos(k_{z})\cos(k_{w}) ].
\end{align}
By diagonalizing the Hamiltonian $H(\mathbf{k})$, we obtain the energy spectrum
\begin{equation}
	E=\pm\sqrt{M(\mathbf{k})^2+\sin^2(k_x)+\sin^2(k_y)+\sin^2(k_z)+\sin^2(k_w)}.
\end{equation}
Each eigenvalue is twofold degenerate, and the spectrum is symmetric with respect to $E=0$.
Therefore, the bulk gap closes when all terms under the square root vanish simultaneously. The corresponding gap-closing conditions are given by $t_1=m/8$, $t_2=(m+2)/16$, $t_2=(2-m)/16$, $t_2=(m-4)/32+3t_{1}/4$, and $t_2=-(m+4)/32-3t_{1}/4$.

In four dimensions, the second Chern number is used to characterize the topological properties of the system~\cite{10.1126/science.294.5543.823}. The second Chern number in 4D momentum space is given by~\cite{PhysRevB.78.195424, 10.1088/2058-9565/aae93b}
\begin{align}
	C_{2}=\frac{1}{4\pi^{2}}\int_{\mathcal{V}} d^{4}\mathbf{k}\,\text{Tr}[\Omega_{xy}\Omega_{zw}+\Omega_{wx}\Omega_{zy}+\Omega_{zx}\Omega_{yw}],
\end{align}
where $\mathcal{V}$ denotes the first Brillouin zone of the 4D momentum space. The non-Abelian Berry curvature is
\begin{align}
	\Omega_{pq}^{\alpha\beta}=\partial_{p}A_{q}^{\alpha\beta}-\partial_{q}A_{p}^{\alpha\beta}+i[A_{p},A_{q}]^{\alpha\beta},
\end{align}
where $p, q=x, y, z, w$, and the Berry connection of the occupied bands is
\begin{align}
	A_{p}^{\alpha\beta}=-i\left\langle \phi^{\alpha}(\mathbf{k})\right|\frac{\partial}{\partial k_{p}}\left|\phi^{\beta}(\mathbf{k})\right\rangle.
\end{align}
Here $\left|\phi^{\alpha}(\mathbf{k})\right\rangle$ denotes the occupied eigenstates below the Fermi energy $E_{F}$ with $\alpha=1, \dots, N_{\rm{occ}}$. We set $N_{\rm{occ}}=2$ in our calculations and employ the adaptive mesh refinement method developed in Ref.~\cite{10.1088/1402-4896/ae8e4d} to evaluate the second Chern number. This method enables accurate and efficient evaluations of $C_2$, particularly near topological phase transition points.

\begin{figure*}[t]
	\includegraphics[width=0.8\textwidth]{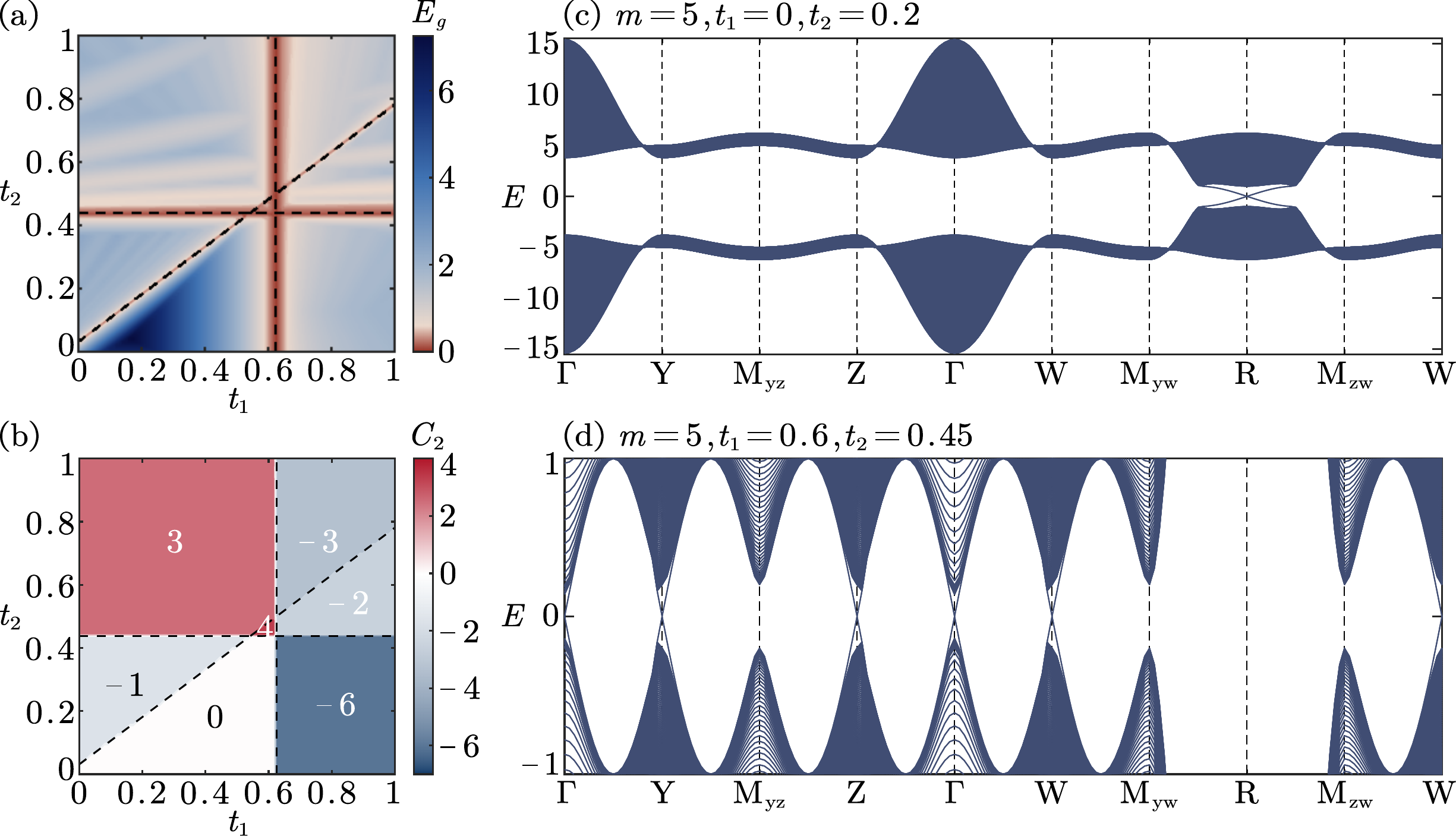} \caption{
		(a) Distribution of the bulk energy gap $E_g$ in the $(t_1,t_2)$ parameter space.
		The color scale represents the magnitude of the bulk energy gap, with the dark-red region corresponding to a nearly closed gap.
		The black dashed lines indicate the gap-closing boundaries determined by $t_1=m/8$, $t_2=(m+2)/16$, and $t_2=(m-4)/32+3t_1/4$, respectively.
		(b) Map of the second Chern number $C_2$ in the $(t_1,t_2)$ parameter space.
		The color scale represents the value of $C_2$, and the numbers in each colored region denote the corresponding second Chern number.
		(c), (d) Energy spectra in the three-dimensional Brillouin zone under open boundary conditions along the $x$ direction for $t_1=0, t_2=0.2$ and $t_1=0.6, t_2=0.45$, respectively.
		Here, $m=5$.
	}%
	\label{fig2}
\end{figure*}

When $t_1=t_2=0$, the Hamiltonian $H(\mathbf{k})$ reduces to the 4D minimal Dirac model. Depending on the value of the Dirac mass, the 4D minimal Dirac model exhibits six topological phases:
\begin{align}
	\begin{split}
		C_{2}(m)=
		\begin{cases}
			0, &m<-4\\
			1, &-4<m<-2\\
			-3, &-2<m<0\\
			3, &0<m<2\\
			-1, &2<m<4\\
			0, &m>4.
		\end{cases}
	\end{split}
\end{align}
When the second Chern number $C_{2}=0$, the system is in a trivial insulator phase. When $C_{2}$ is a nonzero integer, the system is in a topological insulator phase and possesses $|C_{2}|$ three-dimensional gapless boundary states.

\section{Long-range hopping induced emergence of the 4D topological insulating phase}
\label{SecIII}

For $m=5$, the 4D minimal Dirac model is in a trivial insulating phase with a vanishing second Chern number $C_2=0$.
In this section, we investigate the effect of long-range hopping on the 4D trivial insulator.
After introducing the next-nearest-neighbor hopping, we calculate the evolution of the bulk energy gap $E_g$ as a function of the hopping amplitude $t_1$.
As shown in Fig.~\ref{fig1}(a), the bulk gap initially increases with increasing $t_1$, then decreases and closes completely at $t_1=m/8$, followed by a reopening of the gap for larger $t_1$.
This gap-closing and reopening process signals a topological phase transition, which can be characterized by the second Chern number.
Figure~\ref{fig1}(b) shows the evolution of $C_2$ as a function of $t_1$.
For $t_1<m/8$, the system remains in the trivial insulating phase with $C_2=0$.
After the gap closing and reopening, the system undergoes a transition into a 4D topological insulating phase with $C_2=-6$.

We further examine the energy spectra under open boundary conditions along the $x$ direction before and after the topological phase transition.
As shown in Fig.~\ref{fig1}(c), the spectrum remains fully gapped for $t_1=0$, consistent with the trivial insulating phase.
After the transition, gapless boundary modes emerge inside the bulk energy gap, as shown in Fig.~\ref{fig1}(d).
The number of topological boundary modes is consistent with the magnitude of the second Chern number, $|C_2|=6$. Compared with the 4D minimal Dirac model, the topologically nontrivial phase induced by the next-nearest-neighbor hopping is characterized by a high second Chern number.

We further investigate the effects of both next-nearest-neighbor and next-next-nearest-neighbor hoppings on the 4D trivial insulator.
Figure~\ref{fig2}(a) presents the bulk energy gap $E_{g}$ in the $(t_1,t_2)$ parameter space, where the color scale represents the magnitude of the bulk gap and the dark-red regions correspond to nearly gapless points.
The black dashed lines ($t_1=m/8$, $t_2=(m+2)/16$, and $t_2=(m-4)/32+3t_1/4$) denote the gap-closing boundaries obtained by solving the Hamiltonian $H(\mathbf{k})$.
As can be seen, the analytical gap-closing conditions agree well with the bulk gap distribution obtained from numerical diagonalization.

Correspondingly, we calculate the second Chern number in the $(t_1,t_2)$ parameter space, as shown in Fig.~\ref{fig2}(b).
Different colored regions separated by the black dashed lines exhibit distinct second Chern numbers, with the corresponding values labeled inside each region.
These results demonstrate that the combined effects of next-nearest-neighbor and next-next-nearest-neighbor hoppings can drive the 4D trivial insulator into topological insulating phases with various second Chern numbers.

\begin{figure}[t]
	\includegraphics[width=0.48\textwidth]{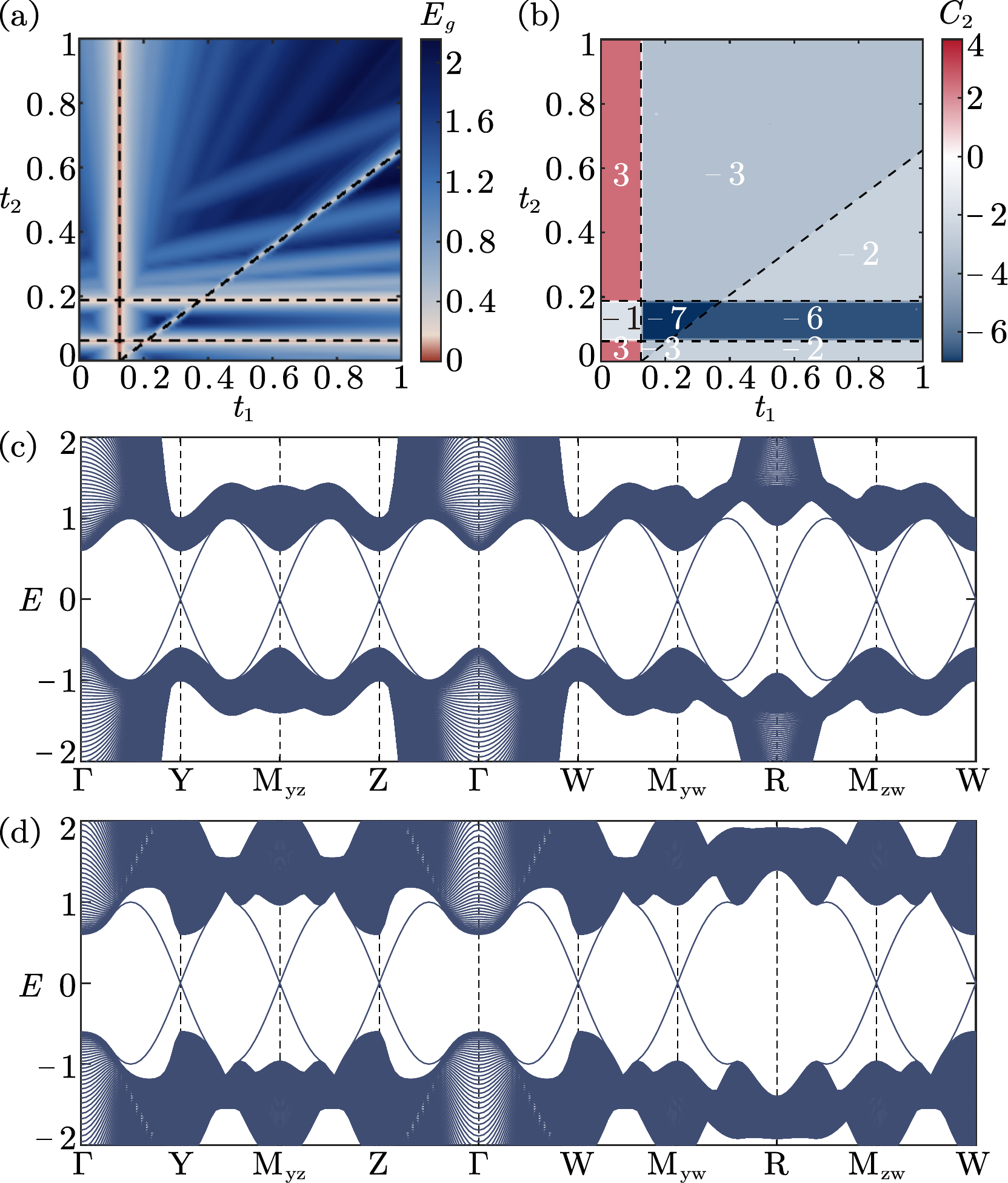} \caption{(a) Distribution of the bulk energy gap $E_g$ in the $(t_1,t_2)$ parameter space when $m=1$.
		The color scale represents the magnitude of the bulk energy gap, with the dark-red region corresponding to a nearly closed gap.
		The black dashed lines indicate the gap-closing boundaries determined by $t_1=m/8$, $t_2=(m+2)/16$, $t_2=(2-m)/16$, and $t_2=(m-4)/32+3t_{1}/4$, respectively.
		(b) Map of the second Chern number $C_2$ in the $(t_1,t_2)$ parameter space when $m=1$.
		The color scale represents the value of $C_2$, and the numbers in each colored region denote the corresponding second Chern number.
		(c), (d) Energy spectra in the three-dimensional Brillouin zone under open boundary conditions along the $x$ direction for $t_1=0.2, t_2=0.15$ and $t_1=0.4, t_2=0.15$, respectively.
		Here, $m=1$.}%
	\label{fig3}
\end{figure}

Even in the presence of only next-next-nearest-neighbor hopping $t_{2}$, the system can be driven from a trivial insulator into a topological insulating phase.
For $m=5$, $t_1=0$, and $t_2=0.2$, the system possesses a second Chern number $C_2=-1$, accompanied by a single gapless three-dimensional boundary mode, as shown in Fig.~\ref{fig2}(c).
Including both next-nearest-neighbor and next-next-nearest-neighbor hoppings gives rise to additional topological phases with even second Chern numbers ($C_2=-2$, $C_2=4$, and $C_2=-6$), which cannot be realized in the 4D minimal Dirac model.

For $m=5$, $t_1=0.6$, and $t_2=0.45$, the system lies in the dark-red triangular region in Fig.~\ref{fig2}(b) with $C_2=4$.
Figure~\ref{fig2}(d) shows the energy spectrum under open boundary conditions along the $x$ direction.
Gapless boundary modes appear around the four high-symmetry points $\Gamma$, $Y$, $Z$, and $W$ in the three-dimensional Brillouin zone, consistent with the nonzero second Chern number.

\section{Tuning the second Chern number of 4D topological insulators}
\label{SecIV}
\begin{figure}[t]
	\includegraphics[width=0.48\textwidth]{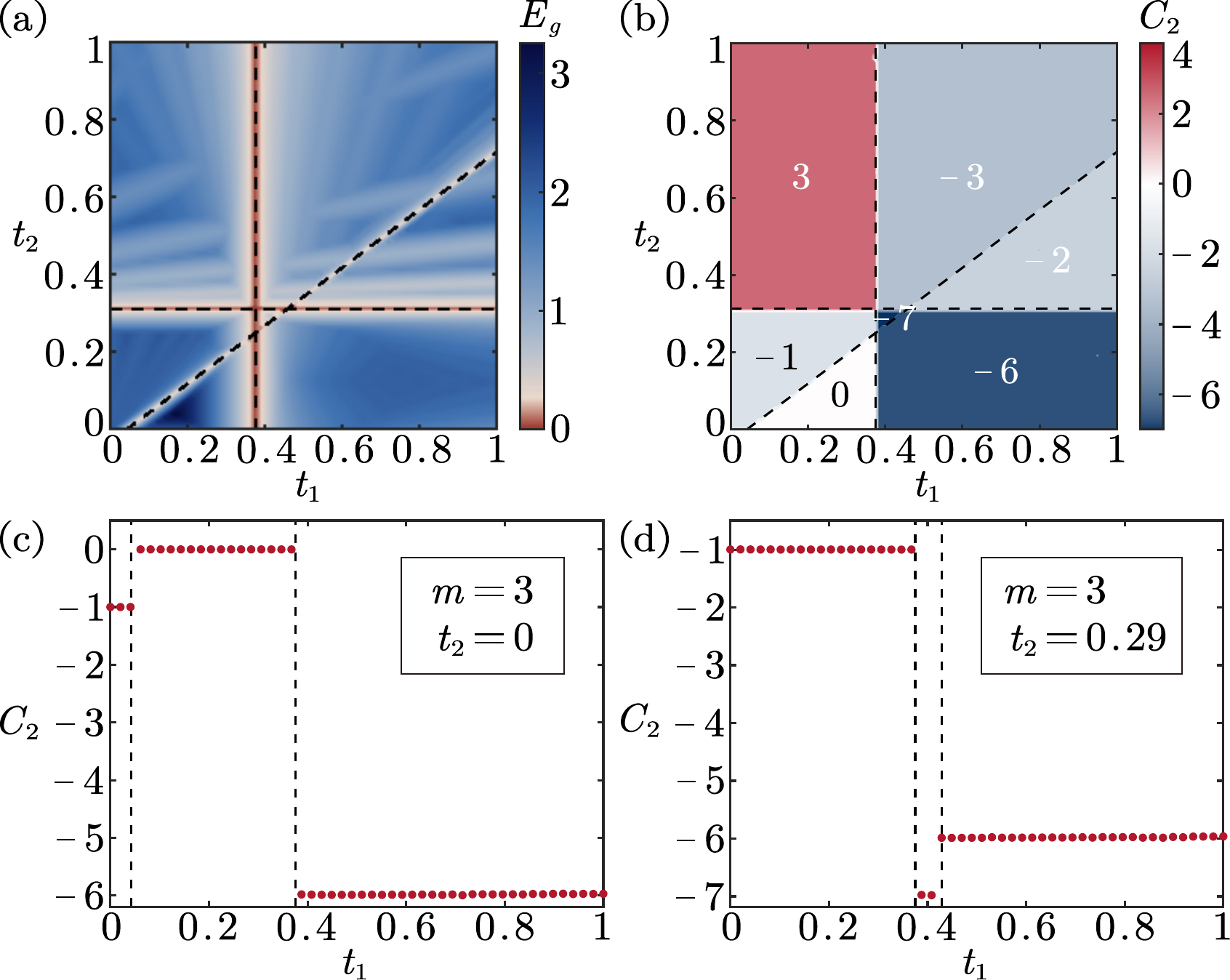} \caption{(a) Distribution of the bulk energy gap $E_g$ in the $(t_1,t_2)$ parameter space when $m=3$.
		The color scale represents the magnitude of the bulk energy gap, with the dark-red region corresponding to a nearly closed gap.
		The black dashed lines indicate the gap-closing boundaries determined by $t_1=m/8$, $t_2=(m+2)/16$, and $t_2=(m-4)/32+3t_{1}/4$, respectively.
		(b) Map of the second Chern number $C_2$ in the $(t_1,t_2)$ parameter space when $m=3$.
		The color scale represents the value of $C_2$, and the numbers in each colored region denote the corresponding second Chern number.
		(c), (d) Second Chern number $C_2$ as a function of $t_1$ for $t_2=0$ and $t_2=0.29$, respectively.
		Here, $m=3$.}%
	\label{fig4}
\end{figure}

In this section, we investigate the effects of long-range hopping on a 4D topological insulator.
First, for $m=1$, the 4D minimal Dirac model is characterized by a nonzero second Chern number $C_2=3$.
Starting from this Hamiltonian, we introduce next-nearest-neighbor and next-next-nearest-neighbor hoppings and study their effects on the topological phases.

Figure~\ref{fig3}(a) shows the bulk energy gap in the $(t_1,t_2)$ parameter space.
The dark-red regions correspond to nearly gapless points.
By solving the Hamiltonian $H(\mathbf{k})$, we obtain the analytical bulk gap-closing boundaries $t_1=m/8$, $t_2=(m+2)/16$, $t_2=(2-m)/16$, and $t_2=(m-4)/32+3t_1/4$, which are indicated by the black dashed lines in Fig.~\ref{fig3}(a).
The regions separated by these gap-closing boundaries remain gapped.

We further calculate the second Chern number in the $(t_1,t_2)$ parameter space, as shown in Fig.~\ref{fig3}(b).
The corresponding values of $C_2$ are labeled in each region for clarity.
When the next-nearest-neighbor hopping is weak, namely $t_1<m/8$, the system exhibits two topological phases with $C_2=3$ and $C_2=-1$, both of which are already present in the 4D minimal Dirac model.
For $t_1>m/8$, the combined effects of next-nearest-neighbor and next-next-nearest-neighbor hoppings lead to a richer topological phase diagram, including new topological phases with the high second Chern number, such as $C_2=-6$ and $C_2=-7$, which cannot be realized in the 4D minimal Dirac model.

\begin{figure}[t]
	\includegraphics[width=0.48\textwidth]{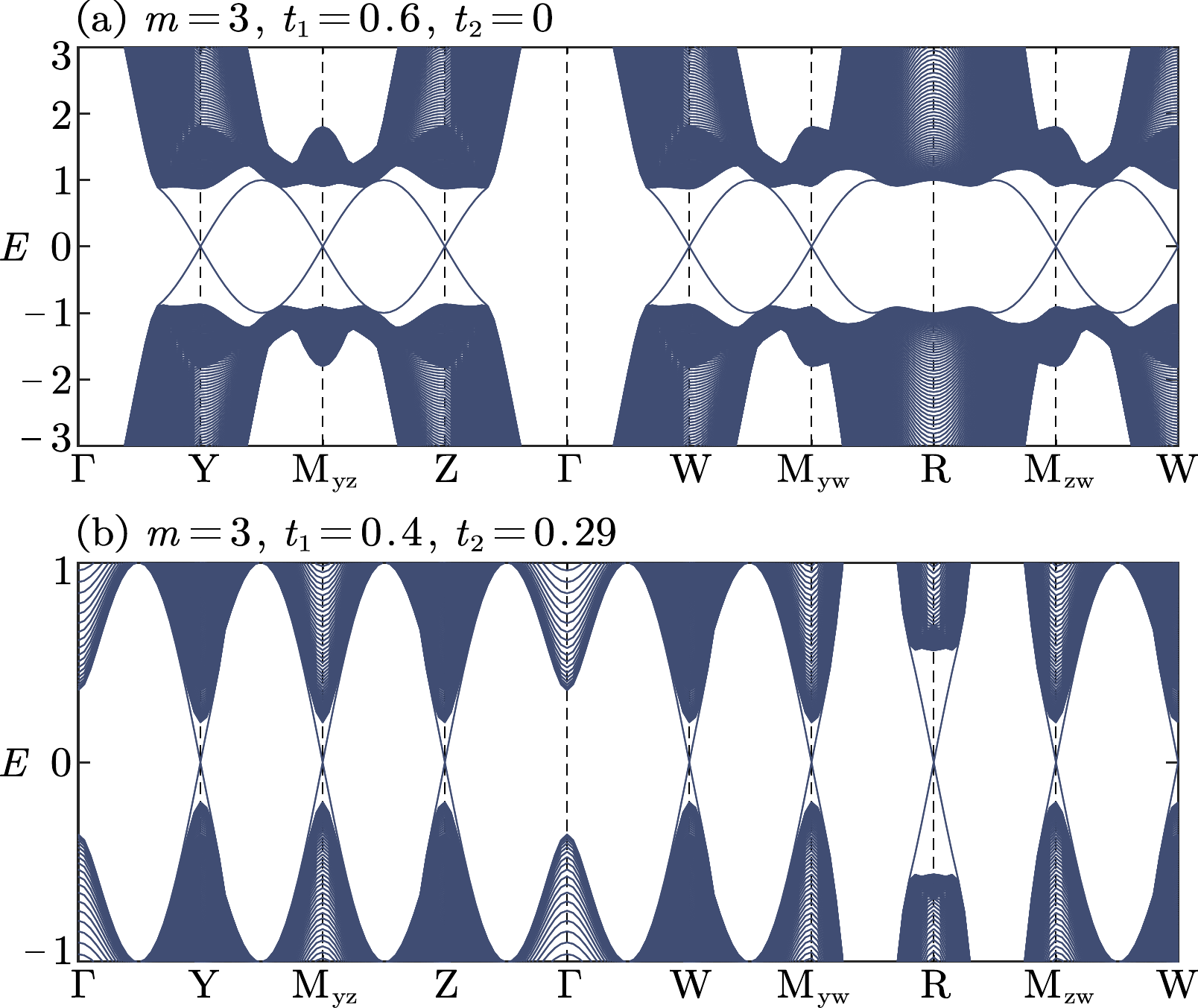} \caption{(a), (b) Energy spectra in the three-dimensional Brillouin zone under open boundary conditions along the $x$ direction for $t_1=0.6, t_2=0$ and $t_1=0.4, t_2=0.29$, respectively.
		Here, $m=3$.}%
	\label{fig5}
\end{figure}

In Figs.~\ref{fig3}(c) and \ref{fig3}(d), we present the energy spectra of the systems with $C_2=-7$ and $C_2=-6$, respectively, under open boundary conditions along the $x$ direction in the three-dimensional Brillouin zone.
Gapless boundary modes emerge inside the bulk energy gap and are located around the high-symmetry points of the three-dimensional Brillouin zone.
The number of topological boundary modes is consistent with the magnitude of the second Chern number $|C_2|$.

Second, for $m=3$, the 4D minimal Dirac model is characterized by a nonzero second Chern number $C_2=-1$.
We investigate the effects of next-nearest-neighbor and next-next-nearest-neighbor hoppings on this topological insulating phase.
Similar to the previous cases, Figure \ref{fig4}(a) shows the bulk energy gap in the $(t_1,t_2)$ parameter space.
The dark-red regions corresponding to nearly closed gaps are consistent with the analytical gap-closing boundaries $t_1=m/8$, $t_2=(m+2)/16$, and $t_2=(m-4)/32+3t_1/4$.

We further calculate the second Chern number in the $(t_1,t_2)$ parameter space, as shown in Fig.~\ref{fig4}(b).
The colors represent different values of $C_2$, and the corresponding values are labeled in each region for clarity.
Along the $t_2=0$ line, as shown in Fig.~\ref{fig4}(c), increasing $t_1$ from zero drives the system from the $C_2=-1$ topological phase into a trivial insulating phase when $t_1=4t_2/3+(4-m)/24$, where the second Chern number vanishes.
Upon further increasing $t_1$ beyond the critical value $t_1=m/8$, the system undergoes another topological transition and enters a topological insulating phase with $C_2=-6$.
In this phase, six gapless three-dimensional boundary modes emerge inside the bulk energy gap, as shown in Fig.~\ref{fig5}(a).

Furthermore, we find that when the next-next-nearest-neighbor hopping amplitude satisfies $(m-1)/8<t_2<(m+2)/16$, increasing $t_1$ drives the system into topological phases with $C_2=-7$ or $C_2=-6$, as shown in Fig.~\ref{fig4}(d).
Figure~\ref{fig5}(b) presents the energy spectra of the $C_2=-7$ phase under open boundary conditions along the $x$ direction.
Seven gapless boundary modes appear inside the bulk energy gap, confirming the bulk-boundary correspondence.

\section{Conclusion}
\label{Conclusion}
In summary, long-range hopping provides an effective mechanism for generating and manipulating high second Chern number phases in 4D systems beyond the minimal Dirac model. By introducing the next-nearest-neighbor hopping and the next-next-nearest-neighbor hopping, a 4D trivial insulator can be transformed into the topological insulator with a high second Chern number $C_2=-6$.
Furthermore, we find that the next-nearest-neighbor hopping and the next-next-nearest-neighbor hopping can drive topological phase transitions in 4D topological insulators, resulting in a rich topological phase diagram that includes topologically nontrivial phases with high second Chern numbers $C_2=-6$ or $C_2=-7$.
The emergence of three-dimensional gapless boundary modes further confirms the bulk-boundary correspondence of these high second Chern number phases. The high second Chern number phases obtained here are expected to exhibit enhanced nonlinear transport responses compared with those in the 4D minimal Dirac model, as the second Chern number directly determines the quantized nonlinear electromagnetic response coefficient in four dimensions~\cite{10.1126/science.294.5543.823}.

Our results establish long-range hopping as a versatile tool for designing and controlling 4D topological states beyond the 4D minimal Dirac model.
Recent advances in artificial quantum systems, including ultracold atoms~\cite{PhysRevLett.115.195303}, electric circuits~\cite{10.1038/s41467-020-15940-3, 10.1093/nsr/nwaa065, 10.1038/s41467-023-36359-6, 10.1038/s41467-023-36767-8, 10.1038/s41467-026-70706-7}, photonic lattices~\cite{PhysRevA.87.013814, PhysRevA.93.043827, 10.1093/nsr/nwac289, 10.1038/nature25011}, and acoustic lattices~\cite{PhysRevX.11.011016}, enable experimental realizations of synthetic 4D topological models.
Furthermore, long-range hopping can be realized in artificial metamaterials such as electric circuits~\cite{PhysRevA.106.012211, PhysRevB.111.014109}, acoustic crystals~\cite{PhysRevApplied.23.024024, PhysRevB.108.205135}, superconducting quantum circuits~\cite{PhysRevB.101.035109}, and superconducting qubits~\cite{PhysRevResearch.3.033288, 10.1126/science.ade7651}. Therefore, we expect that the 4D model with the long-range hopping proposed in this work can be realized in electric circuits and acoustic systems.

\section*{Acknowledgments}
We acknowledge the support of the NSFC (Grants No.~U25D8012, No.~12304195, No.~12504560), the Chutian Scholars Program in Hubei Province, the Hubei Provincial Natural Science Foundation (Grant No.~2025AFA081), the Wuhan city key R\&D program (Grant No. 2025050602030069), the key project of Hubei provincial department of education (Grant No. D20241004), the original seed program of Hubei university.

\bibliographystyle{apsrev4-1-etal-title_6authors}

\end{document}